# Holographic convergent electron beam diffraction (CBED) imaging of two-dimensional crystals

T. Latychevskaia[1,2], S. J. Haigh[3,4], K. S. Novoselov[3,5,6,7]

[1]Institute of Physics, Laboratory for ultrafast microscopy and electron scattering (LUMES), École Polytechnique Fédérale de Lausanne (EPFL) , CH-1015  Lausanne, Switzerland

[2]Paul Scherrer Institute, Forschungsstrasse 111, 5232 Villigen, Switzerland, Switzerland

[3]National Graphene Institute, University of Manchester, Oxford Road, Manchester, M13 9PL, UK

[4]Department of Materials, University of Manchester, Oxford Road, Manchester, M13 9PL, UK

[5]Department of Materials Science and Engineering, National University of Singapore, Singapore, 117575, Singapore

[6]Centre for Advanced 2D Materials, National University of Singapore, 117546 Singapore

[7]Chongqing 2D Materials Institute, Liangjiang New Area, Chongqing, 400714, China

## Abstract

*Convergent beam electron diffraction (CBED) performed on two-dimensional (2D) materials recently emerged as a powerful tool to study structural and stacking defects, adsorbates, atomic 3D displacements in the layers, and the interlayer distances. The formation of the interference patterns in individual CBED spots of 2D crystals can be considered as a hologram, thus the CBED patterns can be directly reconstructed by conventional reconstruction methods adapted from holography. In this study, we review recent results applying CBED to 2D crystals and their heterostructures: holographic CBED on bilayers with the reconstruction of defects and the determination of interlayer distance, CBED on 2D crystal monolayers to reveal adsorbates, and CBED on multilayered van der Waals systems with moiré patterns for local structural determination.*

## 1. Introduction

Convergent beam electron diffraction (CBED) was first demonstrated by Kossel and Möllenstedt in 1939 [1]. Conventionally CBED is performed on three-dimensional (3D) crystals by focussing a convergent electron beam on a small area (about 10 nm in diameter) of a 3D crystal and acquiring the diffraction CBED pattern. Unlike conventional selected area electron diffraction, where the planar illumination produces a diffraction pattern consisting of sharp peaks - in CBED mode the pattern is formed of finite-size disks, whose diameter is determined by the convergence semi-angle of the beam **Fig. 1**. The CBED disks exhibit intensity variations related to the atomic structure and local atomic displacements in the crystal. CBED has been applied for studying crystallographic structure and lattice parameters [2-6], specimen thickness [7], measurements of strain [4, 8], and crystallographic deformations [9, 10]. In large-angle convergent-beam electron diffraction (LACBED) regime, the CBED spots strongly overlap, which allows for an easier and more precise extraction of information on the structure and defects [11, 12]. A good overview of CBED techniques applied for 3D crystal structure determination can be found in references [13-16]. In general, direct interpretation of CBED patterns is not possible and simulations are required to deduce the crystal structure, which often limits wider application of the technique.

Recently, CBED imaging was applied to study van der Waals heterostructures [17-22]. For 2D crystals the interpretation of their CBED patterns is significantly simpler than in the case of 3D crystals because the intensity distributions in individual CBED spots directly map to the atomic arrangement, including local displacements in defects. In CBED the intensity distribution of the scattered wave in the far field can be interpreted as being acquired from a sample that is being illuminated at different angles. This allows capturing of 3D information about atomic positions in the sample. The formation of the interference patterns within the individual CBED spots of 2D crystals can be described by holographic principles, and therefore CBED patterns can be used to directly reconstruct the real space coordinates *via* conventional reconstruction methods adapted from holography. To emphasize the difference in the data analysis we therefore call the technique holographic CBED (HCBED). The reconstructed distributions provide information about the 3D arrangement of atoms in individual layers and in the stack, including the local strain, lattice orientations, local vertical separation between the layers, *etc* which is not accessible by conventional TEM imaging [23, 24]. In this study, we review the recent results in CBED on 2D crystals: holographic CBED on bilayer (BL) with the reconstruction of defects and the measurement of interlayer distance [19, 22], CBED on 2D crystal monolayers (MLs) with imaging of adsorbates [20], and CBED on multilayer van der Waals systems [21].

## 2. Principle of CBED on 2D crystals

### 2.1. CBED spots parameters

In a CBED experiment, the sample *z* position, or defocus $\Delta f$, can be relatively easy changed to move the sample along the optical axis, which allows imaging of the sample with a convergent or divergent electron beam. The CBED arrangement in a convergent wavefront mode ($\Delta f < 0$) is shown in **Fig. 1**. A

CBED pattern from a ML of graphene or hexagonal boron nitride (hBN) consists of finite-sized spots arranged into a six-fold symmetrical pattern. The centres of the spots have the same positions as the corresponding diffraction peaks, defined by the crystal periodicity

$$\sin \vartheta = \lambda / a, \qquad (1)$$

where $\lambda$ is the wavelength, $\vartheta$ is the diffraction angle and $a$ is the period of the crystallographic planes. The diameter of the probing beam on the sample can be evaluated as scaling with the diameter of the limiting aperture as:

$$D = 2|\Delta f| \tan \alpha, \qquad (2)$$

where $\alpha$ is the semi-convergence angle, as shown in **Fig. 1** and where chromatic instabilities and geometric lens aberrations are negligible. For small defocus ($|\Delta f| \approx 0$) the diameter of the probing beam is then given by diffraction on the aperture:

$$D = 1.22 \frac{\lambda}{\tan \alpha}, \qquad (3)$$

where

$$\tan \alpha = \frac{D_A}{2 \Delta f_A}, \qquad (4)$$

$D_A$ is the limiting aperture diameter and $\Delta f_A$ is the distance between the limiting aperture and the virtual source plane, as shown in **Fig. 1**. By changing the *z* position of the sample in the convergent electron beam, the defocus $\Delta f$ is changed, which in turn changes the diameter of the probing electron beam according Eqs. (2) and (3). The diameters of the CBED spots however do not change, because they are defined by the limiting aperture angular size or tan α, as given by Eq. (4). For example, the diameter of the zero-order CBED spot size can be derived from geometrical considerations as shown in **Fig. 1**. Thus, the semi convergence angle α, or the diameter of the limiting aperture, defines the diameter of the CBED spots and for HCBED it should be selected such that the CBED spots from a single crystal structure do not overlap. Then, the size of the probing beam, or the probed area, can be regulated by selecting $\Delta f$. Note that this is equivalent to adjusting the sample *z* position where lens aberrations are neglected.

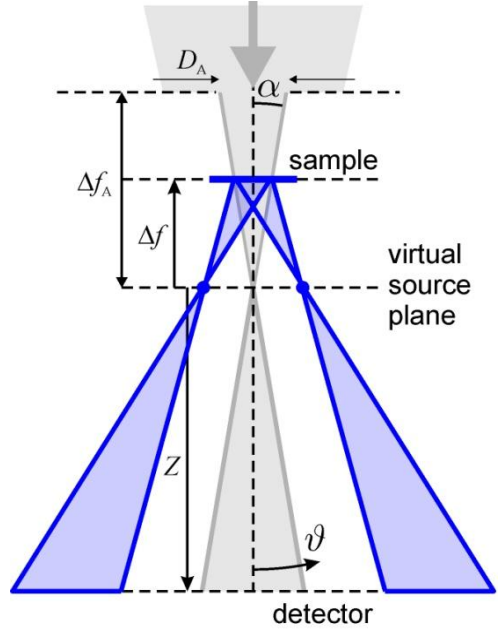

**Fig. 1**. CBED arrangement.

## 2.2. Probing wavefront distribution

Where the probing wavefront is formed by a diffraction on the limiting aperture, the intensity distribution is described by Fresnel diffraction on a round aperture:

$$\psi_0(\vec{r}) \propto \iint a(\vec{r}_0) \frac{\exp(-ikr_0)}{r_0} \frac{\exp(ik|\vec{r}_0 - \vec{r}|)}{|\vec{r}_0 - \vec{r}|} d\vec{r}_0, \qquad (5)$$

where $a(\vec{r}_0)$ is the aperture function, $\vec{r}_0 = (x_0, y_0, z_0)$ is the coordinate in the aperture plane and $\vec{r} = (x, y, z)$ is the coordinate in the sample plane. For defocus values $\Delta f \approx 0$, the probing wavefront distribution is described by Fraunhofer diffraction on a round aperture, that is, by Fourier transform (FT) of the aperture distribution. Examples of probing wavefront distributions are shown in **Fig. 2**.

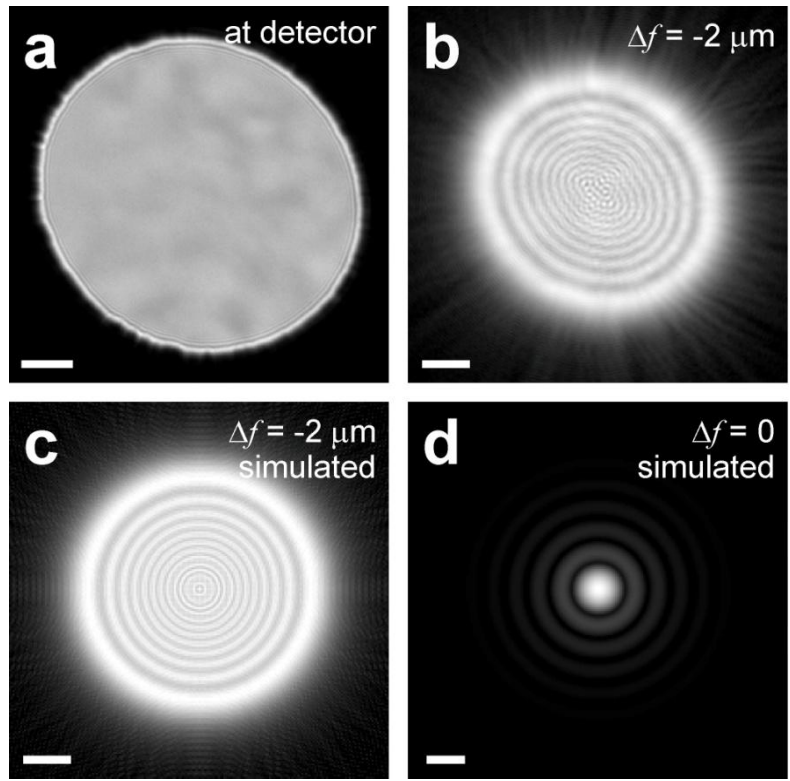

**Fig. 2**. Experimental and simulated images of the probing wave at different focal planes in the transmission electron microscope. (a) Experimentally measured intensity distribution in the zero-order CBED spot at $\Delta f = -2$ μm. (b) Amplitude of the probing wavefront at $\Delta f = -2$ μm, obtained by backward propagation of the wavefront from the detector plane (shown in (a)). (c) Amplitude of the simulated wavefront at $\Delta f = -2$ μm. (d) Amplitude of the simulated wavefront at $\Delta f = 0$. The scalebars correspond to (a) 100 nm, (b) and (c) to 5 nm, and (d) to 0.5 nm. The experimental images were obtained with Titan ChemiSTEM microscope operated at 80 kV, with an ultra-stable high-brightness Schottky FEG source. Note that the experimental data shows evidence of a slight astigmatism. Adapted from [19].

## 3. CBED on 2D crystal monolayers

CBED on 2D crystal MLs is studied in detail in reference [20], and here we present the main results from the study.

### 3.1. In-plane and out-of plane ripples

### 3.1.1 Phase shifts caused by atomic displacements, geometrical approach

A 2D ML crystal is a perfect test sample to explain the basics of the intensity distributions observed in individual CBED spots. Atomic misalignments in the form of out-of-plane and in-plane shifts are illustrated in **Fig. 3**.

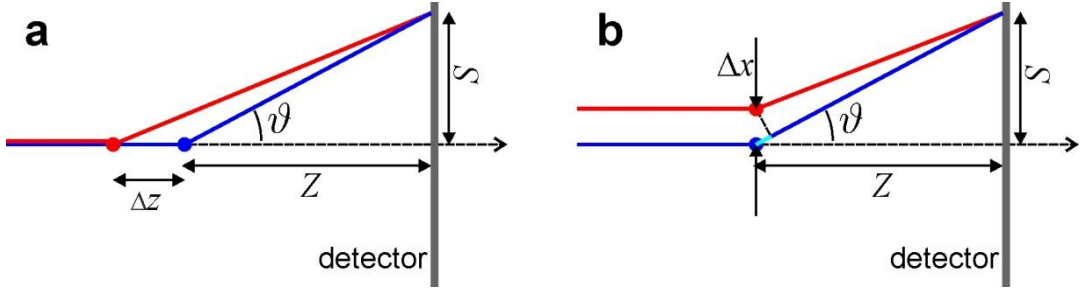

**Fig. 3**. Geometrical arrangement of scattering from two atoms separated by (a) $\Delta z$ and (b) $\Delta x$ distances. (a) Where two atoms are separated by $\Delta z$, the path difference is given by $\Delta l = l_{\text{blue}} - l_{\text{red}} \approx \Delta z (1 - \cos \vartheta)$. (b) Where two atoms are separated by $\Delta x$, the optical path difference (shown in cyan) is given by $\Delta l = \Delta x \sin \vartheta$.

From the geometrical considerations, shown in **Fig. 3**, the phase differences due to atomic shifts are:

$$\Delta \varphi_z = \frac{2\pi}{\lambda} \Delta z (1 - \cos \vartheta) \quad (6)$$

and

$$\Delta \varphi_x = \frac{2\pi}{\lambda} \Delta x \sin \vartheta \quad (7)$$

for out-of-plane $\Delta z$ and in-plane $\Delta x$ atomic shifts, respectively. From Eqs. (6) and (7) we see that atomic $\Delta x$ shifts cause stronger phase shifts because $\Delta \varphi_x \propto \vartheta$ while $\Delta \varphi_z \propto \vartheta^2$. Thus, in-plane shifts introduce stronger intensity variations into HCBED spots interference patterns, and therefore they are easier to detect and reconstruct.

### 3.1.2 Phase shifts caused by atomic displacements, wave theory approach

Alternatively, using optical wave theory the phase shifts of the waves scattered from misaligned atoms can be derived as follows. The distribution of the scattered wave in the far-field $U(\vec{R})$ is given by:

$$U(\vec{R}) \approx \int \frac{\exp(-ikr)}{r} t(\vec{r}) \frac{\exp(ik|\vec{r}-\vec{R}|)}{|\vec{r}-\vec{R}|} d\vec{r} \propto \frac{\exp(ikR)}{R} \int \frac{\exp(-ikr)}{r} t(\vec{r}) \exp\left(-ik\frac{\vec{r}\vec{R}}{R}\right) d\vec{r}, \quad (8)$$

where we consider a convergent probe wavefront $\frac{\exp(-ikr)}{r}$, $t(\vec{r})$ is the transmission function of the sample, $\vec{R}=(X,Y,Z)$ is the coordinate in the detector plane, and using $R \gg r$ the approximation $|\vec{r}-\vec{R}| \approx R - \frac{\vec{r}\vec{R}}{R}$ (9)

is applied. We introduce *K*-coordinates to simply the algebraic expression:

$$\vec{K} = (K_x, K_y, K_z) = k\frac{\vec{R}}{R} = \frac{2\pi}{\lambda R}(X,Y,Z),\ |\vec{K}| = k = \frac{2\pi}{\lambda},\ K_z = \sqrt{K^2 - K_x^2 - K_y^2} \quad (10)$$

and re-write:

$$U(K_x, K_y) \approx \exp(ikR)\int \exp(-ikr)t(x,y,z)\exp\left[-i(K_x x + K_y y)\right]\exp(-izK_z)\mathrm{d}x\mathrm{d}y\mathrm{d}z. \quad (11)$$

The wavefront scattered by an atom at $\vec{r}=(x,y,z)$ is given by:

$$U(K_x, K_y) \propto \exp(ikR)\exp(-ikr)\exp\left[-i(xK_x + yK_y)\right]\exp(-izK_z). \quad (12)$$

*3.1.2.1 Phase shift caused by an out-of-plane displacement*

Wavefronts scattered by atoms positioned at $\vec{r}_1 = (0,0,-|\Delta f|)$, $r_1 = |\Delta f|$ and $\vec{r}_2 = (0,0,-|\Delta f|+\Delta z)$, $r_2 = |-|\Delta f|+\Delta z|$, are given by:

$$\begin{aligned}U_1(K_x, K_y) &\propto \exp(ikR)\exp(-ik|\Delta f|)\exp\left[iK_z|\Delta f|\right]\\ U_2(K_x, K_y) &\propto \exp(ikR)\exp\left[ik(-|\Delta f|+\Delta z)\right]\exp\left[-iK_z(-|\Delta f|+\Delta z)\right].\end{aligned} \quad (13)$$

The corresponding phases of the wavefronts are:

$$\begin{aligned}\varphi_1 &= kR - k|\Delta f| + \Delta f K_z\\ \varphi_2 &= kR + k(-|\Delta f|+\Delta z) - (-|\Delta f|+\Delta z)K_z,\end{aligned} \quad (14)$$

and the phase difference is given by:

$$\Delta\varphi_z = \varphi_2 - \varphi_1 = \Delta z(k - K_z) = \Delta z\frac{2\pi}{\lambda}(1-\cos\vartheta), \quad (15)$$

where we applied $K_z = \frac{2\pi}{\lambda}\cos\vartheta$. The obtained phase shift agrees with the phase shift derived from geometrical consideration, Eq. (6) and Fig. 3(a). The CBED pattern and phase shifts produced by a ML with out-of-plane atomic shifts forming a "bulge" are shown in **Fig. 4(a)-(c)**.

### 3.1.2.2 Phase shift caused by an in-plane displacement

Wavefronts scattered by two adjacent atoms positioned at $\vec{r}_1 = (0, 0, -|\Delta f|)$ and $\vec{r}_2 = (a + \Delta x, 0, -|\Delta f|)$, where $r_1 = |\Delta f|$ and $r_2 = \sqrt{(\Delta f)^2 + (a + \Delta x)^2} \approx |\Delta f|$, and $a$ is the lattice period, are given by:

$$U_1(K_x, K_y) \propto \exp(ikR)\exp(-ik|\Delta f|)\exp[iK_z|\Delta f|]$$
$$U_2(K_x, K_y) \propto \exp(ikR)\exp(-ik|\Delta f|)\exp[-iK_x(a + \Delta x)]\exp(iK_z|\Delta f|). \quad (16)$$

The corresponding phases of the wavefronts are:

$$\varphi_1(K_x, K_y) = kR - k|\Delta f| + K_z|\Delta f|$$
$$\varphi_2(K_x, K_y) = kR - k|\Delta f| - K_x(a + \Delta x) + K_z|\Delta f|, \quad (17)$$

and the phase difference is given by:

$$\Delta\varphi_x(K_x, K_y) = \varphi_2(K_x, K_y) - \varphi_1(K_x, K_y) = -K_x(a + \Delta x). \quad (18)$$

When $\Delta x = 0$, the phase shift $\Delta\varphi = 2\pi$, and we obtain:

$$\Delta\varphi_x(K_x, K_y) = -K_x^{(1)} a = 2\pi, \quad (19)$$

which corresponds to the position of the *n*-th-order diffraction peak:

$$K_x^{(n)} = n\frac{2\pi}{a}. \quad (20)$$

The phase shift due to a lateral shift $\Delta x$ is given by:

$$\Delta\varphi_x(K_x, K_y) = -K_x \Delta x = -\frac{2\pi}{\lambda}\Delta x \sin\vartheta \quad (21)$$

which is an odd function of $K_x$. Thus, for $\Delta x \neq 0$ there will be an additional phase shift in opposite CBED spots, as for example in the spots $(\bar{1}010)$ and $(10\bar{1}0)$, and these phase shifts will be of opposite sign. The obtained phase shift agrees with the phase shift derived from geometrical considerations, Eq. (7) and **Fig. 3(b)**. The CBED pattern and phase shifts produced by a ML with in-plane atomic shifts forming a linear lateral shift is illustrated in **Fig. 4(d)-(f)**.

The results shown in **Fig. 4(b)** and **(e)** demonstrate that it is easy to distinguish between out-of-plane and in-plane atomic displacements even without performing a reconstruction simply by comparing the intensity contrast in opposite CBED spots: An out-of-plane defect will always result in symmetric variations in intensity distribution between mirror-symmetric CBED spots, and an in-plane defect will result in antisymmetric variations.

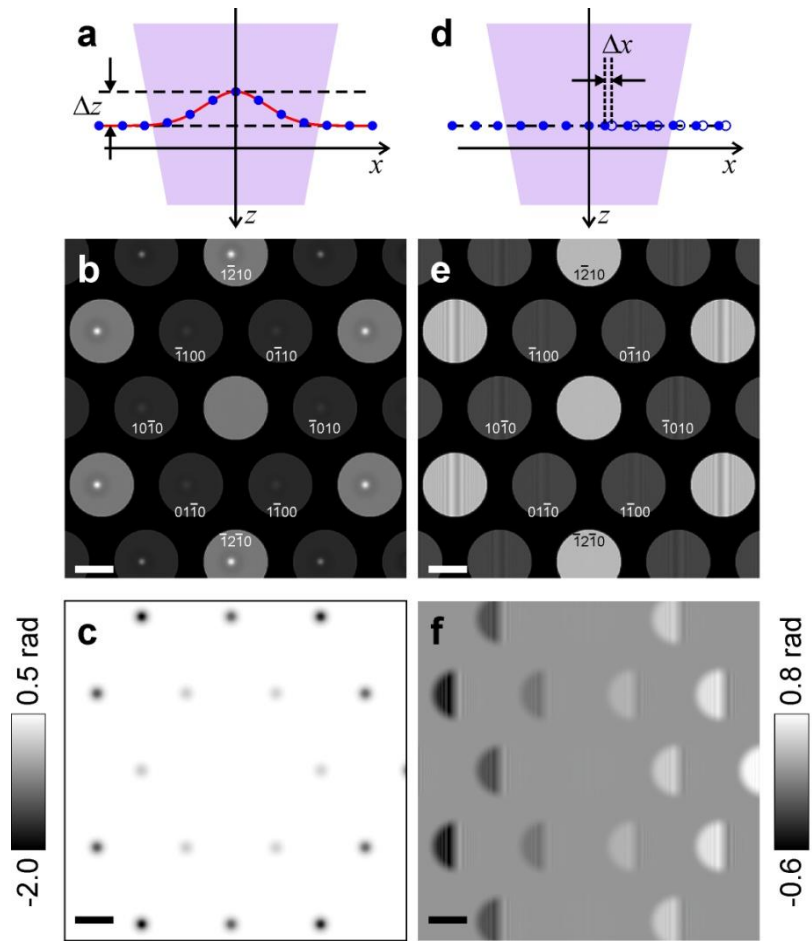

**Fig. 4**. CBED patterns of a graphene monolayer with (a) - (c) an out-of-plane "bulge" and (d) - (f) an in-plane lateral displacement both simulated for $\Delta f$ = -2μm (underfocus). (a) Sketch of the side view of a graphene layer with atoms displaced out of plane in form of a "bulge". (b) Corresponding simulated CBED pattern, the "bulge" height is $|\Delta z|$ = 2 nm. (c) Phase shift introduced by the lattice deformation into the probing electron wave, calculated at the detector plane. (d) Sketch of the side view of a graphene layer with atoms displaced in plane in form of a lateral shift. (e) The corresponding simulated CBED pattern, where the atoms positioned at $x > 0$ are displaced by $\Delta x$ = -10 pm. (f) Phase shift introduced by the lattice deformation into the probing electron wave, calculated at the detector plane. For these simulations the probed area is about 30 nm in diameter. The scale bars in (b), (c), (e) and (f) correspond to 2 nm$^{-1}$. Adapted from [20].

### 3.2. Imaging adsorbates on MLs

Lattice deformations such as strain or rippling do not cause noticeable intensity variations in the zero-order CBED spot, and cause increasing intensity variations as the order of the CBED spot increases (spots are further from the zero-order spot and correspond to higher spatial frequencies). In contrast, adsorbates exhibit strong intensity variations in all CBED spots. The zero-order CBED spot is in fact an in-line hologram of the sample, and in-line holography is known to exhibit high intensity contrast that is very valuable for imaging of weak phase objects that are not detectable for conventional TEM at Gaussian focus. The zero-order CBED spot of a ML with phase adsorbates on one surface displays an inversion of contrast when acquired at $\Delta f > 0$ and $\Delta f < 0$, defocus values respectively [20]. In the case of an adsorbate which is not a phase object so also absorbs electrons - the zero-order CBED spot displays a dark feature.

Experimental CBED patterns of an hBN sample acquired at $\Delta f = 5$ μm and $\Delta f = -5$ μm are shown in **Fig. 5**. The corresponding high angle annular dark field (HAADF) scanning transmission electron microscope (STEM) images acquired before and after CBED imaging demonstrate no visible radiation damage to the sample (**Fig. 5(a)** and **(b)**). CBED patterns acquired at $\Delta f > 0$ and $\Delta f < 0$ (**Fig. 5(c)** and **(e)** respectively) show inversion of the contrast, as expected for weak phase objects [20]. In addition, the intensity distribution within a selected CBED spot demonstrates centro-symmetrical flipping upon transition from $\Delta f > 0$ and $\Delta f < 0$ (see the magnified zero-order CBED spots shown in **Fig. 5(d)** and **(f)**). The zero-order CBED spot is an in-line hologram of the probed sample region formed by interference of the scattered and non-scattered waves. The higher-order CBED spots are formed only by the scattered wave and therefore show different intensity distributions than the zero-order CBED spot. The higher-order CBED spots are highly sensitive to the atomic misalignments as discussed above, and their intensity distribution reflects the ripples in the 2D crystal. For example, in the first- and second-order CBED spots there is a darker feature as indicated by magenta arrow in **Fig. 5(f), (g)** and **(h)**. Such a dark feature is not observed in the zero-order CBED spot and thus cannot be attributed to an adsorbate. The HAADF STEM images of the sample also confirm that there are no adsorbates at this location, as indicated by the magenta arrows in **Fig. 5(a)** and **(b)**. Thus, the dark feature can be explained by a surface deformation in that location, probably because of the stress between the two adjacent adsorbates.

The zero-order CBED spot can be treated as an in-line hologram, and the amplitude and phase distributions of the transmission function of the sample can be reconstructed by available algorithms [25]. An example of such reconstruction is shown in **Fig. 6**. Adsorbate patches of sub-nanometre sizes are observed in the reconstructed phase distributions and can be cross-validated with the corresponding HAADF STEM images. Also here, the reconstructed amplitude distributions exhibit blurred structure when compared to the reconstructed phase distributions, which suggests that the imaged objects are phase objects without any absorption contrast.

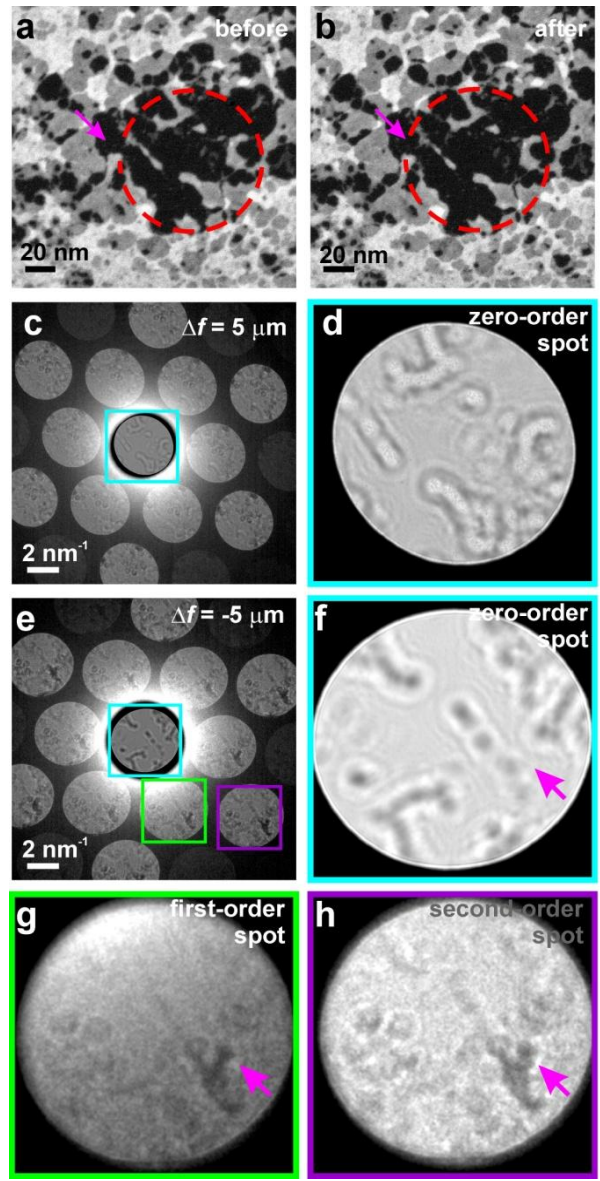

**Fig. 5**. CBED patterns of ML hBN acquired at $\Delta f > 0$ and $\Delta f < 0$.

(a) – (b) High angle annular dark field (HAADF) scanning transmission electron microscope (STEM) images of the sample before and after CBED imaging; the imaged area is marked by the red dashed circle.

(c) CBED pattern acquired at $\Delta f$ = 5 µm and (d) the intensity distribution in the zero-order CBED spot.

(e) CBED pattern acquired at $\Delta f$ = - 5 µm and magnified images of the intensity distributions of (f) the zero-order, (g) the first-order, and (h) the second-order CBED spots. Adapted from [20].

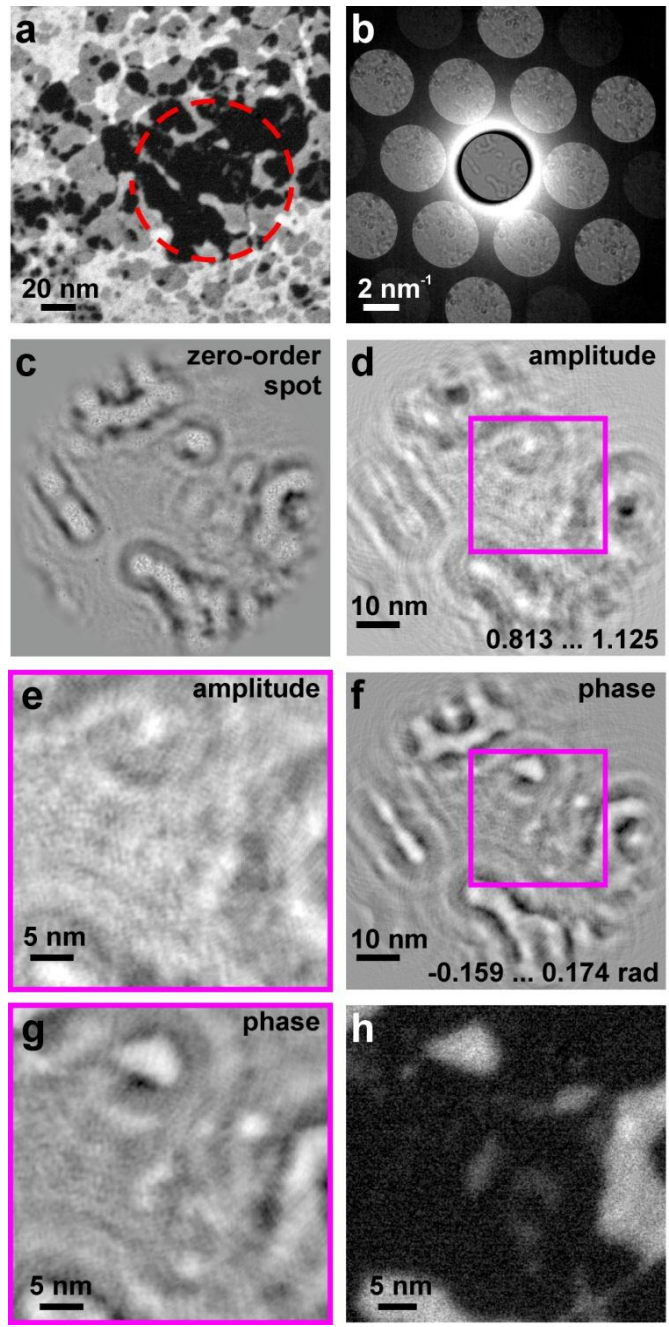

**Fig. 6.** Reconstruction of zero-order CBED spot as in-line hologram. (a) High angle annular dark field (HAADF) images of the sample before CBED imaging; the imaged area is marked by the red dashed circle. (b) CBED pattern. (c) Zero-order CBED spot. (d) – (e) The amplitude distribution reconstructed from the zero-order CBED spot at $\Delta f$ = 5 μm. (f) – (g) The phase distribution reconstructed from the zero-order CBED spot at $\Delta f$ = 5 μm. (h) Zoomed-in region in the HAADF image for comparison with the reconstructed amplitude and phase distributions of the same region (marked by the magenta square). The vales given at the bottom of (d) and (f) are the range of the reconstructed amplitude and phase values, respectively. Adapted from [20].

# 4. Holographic CBED of BLs

## 4.1. Formation of interference pattern in CBED spots

For CBED of a BL sample, the two sets of electron beams diffracted on each layer interfere in the detector plane, creating interference patterns at the positions of the overlapping CBED spots, as illustrated in **Fig. 7**. Such interference patterns contain rich information about the local inter-atomic spacing (local strain), the vertical distance between the layers, the relative orientation between the layers, etc. The period and tilt of the fringes in the interference pattern can be explained by considering the position of the sources in the virtual source plane, as sketched in **Fig. 7(a)**. A real-space illustration of a BL sample with an in-plane twist angle $\beta$ is shown in **Fig. 7(b)**. In the virtual source plane, the Bragg diffraction peaks create virtual sources, which are rotated by the twist angle $\beta$ relative to each other, as depicted in **Fig. 7(c)**.

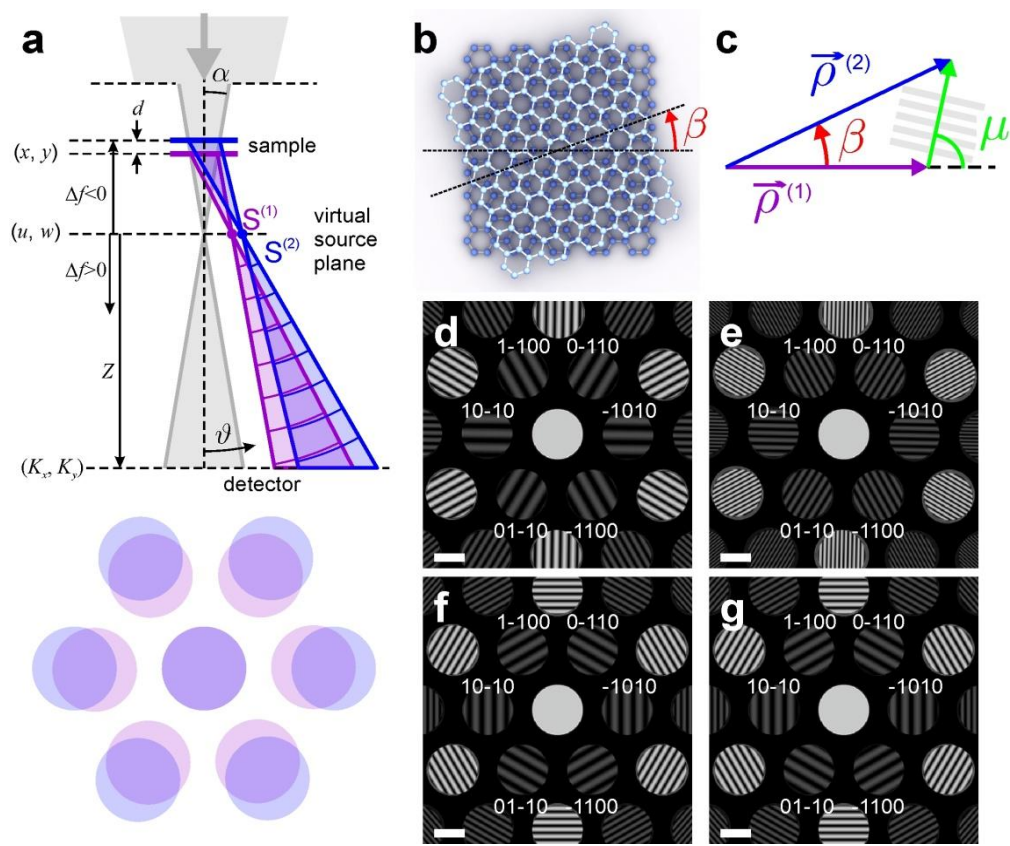

**Fig. 7**. CBED on a bilayer system. (a) Top: Schematics of the HCBED experimental arrangement. $\Delta f$ is the sample $z$-position counted from the focus of the electron beam (in this particular case underfocus, $\Delta f < 0$, CBED conditions are shown), $Z$ is the distance from the virtual source plane to the detector, $S^{(1)}$ and $S^{(2)}$ are the virtual sources for the first-order CBED spots of bottom (1) and top (2) crystals in the heterostructure stack

respectively. $\vartheta$ is the angular coordinate on the detector. Bottom: Distribution of CBED spots on a detector in the case of two aligned crystals (no relative twist $\beta = 0$), but with slightly different lattice constants. (b) Real-space distribution of a BL sample with a difference in lattice parameter *and* a relative in-plane twist angle $\beta$. (c) Arrangement of the vectors in the virtual source plane for the situation shown in (b). (d) - (g) Simulated CBED patterns for: (d) and (e) BL graphene in AA stacking with twist angle 1° and 2°, respectively; (f) and (g) graphene-hBN BL stacking without twist, in AA and AB stacking configurations, respectively. The scale bars in (d) - (g) correspond to 2 nm$^{-1}$.

The interference pattern in a CBED spot is described by the formula [22]:

$$I(\vec{K}) \propto 2 + 2\cos\left\{\vec{K}\Delta\vec{\rho} + \frac{\pi}{\lambda}\left[\frac{\left(\rho^{(2)}\right)^2}{|\Delta f + d|} - \frac{\left(\rho^{(1)}\right)^2}{|\Delta f|}\right] - \Delta\gamma\right\}, \tag{22}$$

where $\Delta\vec{\rho} = \vec{\rho}^{(1)} - \vec{\rho}^{(2)}$ and $\Delta\gamma = \gamma^{(1)} - \gamma^{(2)}$, $\vec{\rho}^{(1)}$ and $\vec{\rho}^{(2)}$ are the positions of the virtual sources in the virtual source plane, while $\gamma_m^{(1)}$ and $\gamma_m^{(2)}$ are the constant phases of the virtual sources. The first term in the argument of the cosine term describes the interference fringes distribution. The period of the interference fringes is given by:

$$T = \frac{2\pi}{\Delta\rho}. \tag{23}$$

The tilt of the interference fringes $\mu$ can be found from the geometrical arrangement of the vectors in the virtual source plane (**Fig. 7(c)**):

$$\tan\mu = \frac{\rho^{(2)}\sin\beta}{\rho^{(2)}\cos\beta - \rho^{(1)}}. \tag{24}$$

The higher the twist angle, the larger $\Delta\rho$ and the smaller is the period of the fringes, as can be seen in the simulations of BL graphene at twist angles of 1° and 2° in **Fig. 7(d)** and **(e)**, respectively. The second term in the argument of cosine in Eq. (22) is a constant offset. The third term $\Delta\gamma$ depends on the local stacking of the layers (for example, AA or AB crystal stacking) and defines the position of the centre of the interference pattern. If the local stacking under the centre of the electron beam is AA, then $\Delta\gamma = 0$ and the interference pattern within an overlapping CBED spot has a maximum at the centre of the CBED spot. For AB stacking $\Delta\gamma \neq 0$ and the interference pattern within an overlapping CBED spot does not display a maximum at the centre of the CBED spot, as can be seen in the simulations of a graphene-hBN system for AA and AB stacking in **Fig. 7(f)** and **(g)**, respectively.

## 4.2 Holographic reconstruction

### 4.2.1 Protocol of the reconstruction procedure

The atomic displacements relatively to their position in perfect lattices for a BL system can be reconstructed from the corresponding CBED pattern when treating the CBED spots as off-axis holograms. The reconstruction procedure is described in references [19, 22]. Here we provide the main reconstruction steps, also illustrated in **Fig. 8**:

(1) For holographic reconstruction, a CBED spot is selected with the centre at an arithmetic average of the centres of the individual CBED spots.

(2) The 2D Fourier spectrum of the selected region is calculated. In the obtained complex-valued spectrum, one zero-order and two sidebands are observed (**Figs. 8(a)** and **(b)**).

(3) One of the sidebands is selected while the zero-order and remaining sideband are set to zero (**Fig. 8(b)**). Note, it is crucially important that the sidebands in different CBED spots are chosen consistently. The chosen sidebands in different CBED spots have to be related by the same symmetry transformations as the CBED spots themselves (rotation by an integer number of $\pi/3$).

(4) The whole spectrum is shifted so that the maximum of the sideband is located at the origin of the reciprocal plane (**Fig. 8(b)**).

(5) The inverse 2D Fourier transform the resulting distribution is calculated.

(6) The amplitude and phase distributions are extracted from the obtained complex-valued distribution, (**Fig. 8(c)**).

In steps (3) and (4), the right sideband is selected at the position defined by the fringes tilt angle $\mu$ and period $T$, as defined by Eqs. (23) and (24) and explained in reference [22].

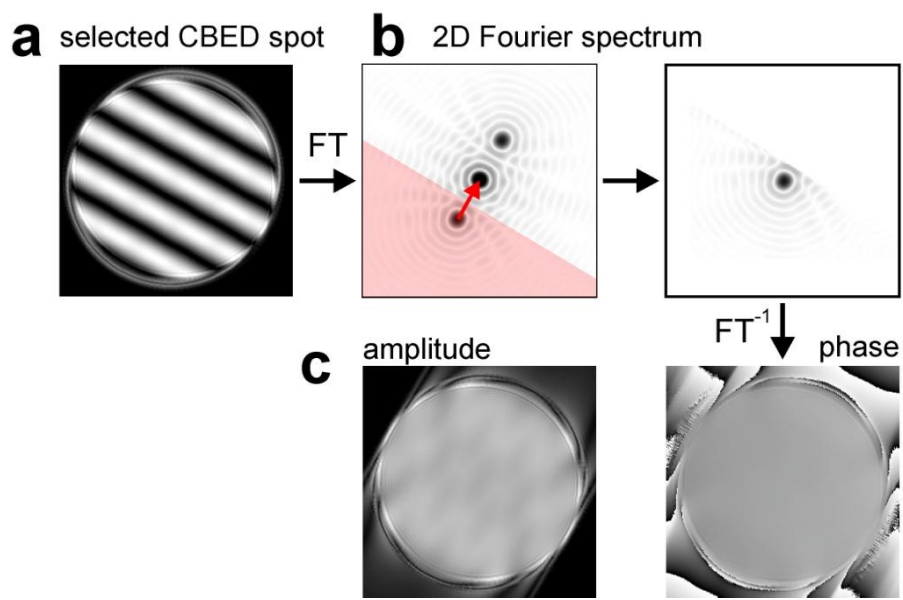

**Fig. 8**. Reconstruction of a CBED spots as off-axis holograms. (a) Selected CBED spot. (b) Amplitude of its Fourier spectrum. The area shaded red is selected and shifted so that the sideband peak is in the centre. (c) Inverse 2D FT gives the complex-valued distribution, where the amplitude and phase are extracted.

After the amplitude and phase distributions for each CBED spot are reconstructed, only the phase distributions are considered since only these carry the information about the atomic positions. The individual reconstructed phase distributions are averaged, that is, all six distributions are added together and divided by six, and the out-of-plane atomic shifts are calculated by Eq. (15). The in-plane atomic shifts are then calculated from the reconstructed phases of the opposite CBED spots by applying Eq. (21).

### 4.2.2 Simulated example of reconstruction in-plane and out-of-plane displacement

An example of the simulated and reconstructed CBED pattern for a graphene-hBN system with both out-of-plane and in-plane shifts is shown in **Fig. 9**.

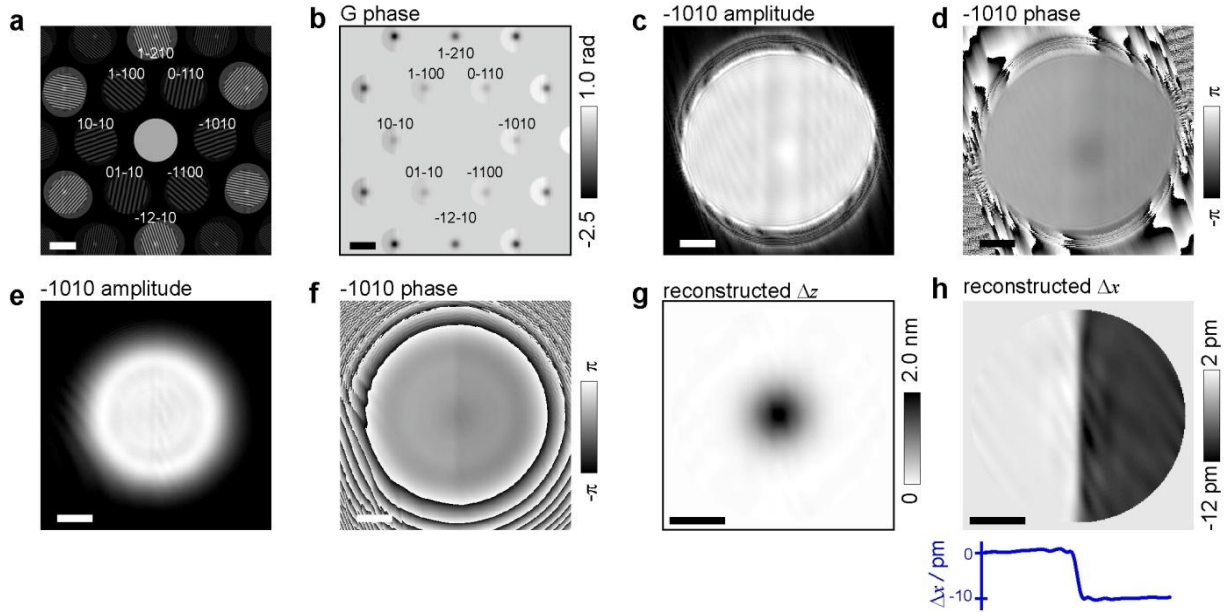

**Fig. 9**. Simulated CBED pattern of a graphene-hBN bilayer where the graphene lattice is deformed, the twist angle 4°, and the reconstructed in-plane and out-of-plane graphene lattice deformations.
(a) Simulated CBED pattern where the atoms in the graphene layer are mis-positioned as follows: the atoms positioned at $x > 0$ are displaced by $\Delta x = -10$ pm, and the atomic $z$-positions are shifted by $\Delta z = -A_B \exp\left(-\dfrac{x^2 + y^2}{2\sigma_B^2}\right)$, $A_B$ = 2 nm, $\sigma_B$ = 2 nm. For these simulations $\Delta f$ = -2 µm, the distance between the layers is 3.35 Å, the probed area is about 28 nm in diameter.
(b) The difference of the phases of the wavefronts scattered by the graphene layer with and without atomic displacement, calculated as $\Delta\varphi = \varphi_{\text{with}} - \varphi_{\text{without}}$; only the phase difference is shown, the phase distribution $\varphi_{without}$ (not shown) exhibits constant values.
(c) Amplitude of the reconstructed wavefront at the (-1010) CBED spot.
(d) Phase of the reconstructed wavefront at the (-1010) CBED spot.
(e) Amplitude of the reconstructed complex-valued distribution from the (-1010) CBED spot after correction for defocus.
(f) Phase of the reconstructed complex-valued distribution from the (-1010) CBED spot after correction for defocus.
(g) Reconstructed distribution of the atomic out-of-plane displacement, $\Delta z$, in the graphene layer.
(h) Reconstructed distribution of the atomic in-plane displacement, $\Delta x$, in the graphene layer.
The scale bars in (a) and (b) correspond to 2 nm$^{-1}$. The scale bars in (c) – (h) correspond to 5 nm. Adapted from [19].

### 4.2.3 Experimental example of reconstruction in-plane and out-of-plane displacement

An experimental CBED pattern from a twisted BL system (graphene-hBN) is shown in **Fig. 10.** CBED spots originating from the graphene layer are found at a slightly larger diffraction angle than CBED spots from the hBN layer (due to differences in the in plane lattice parameter), which allows an easy assignment of the CBED spots corresponding to different layers. A stacking fault between the layers is evident by the presence of a distinctive ridge in the interference patterns in the first- and higher-order CBED spots (**Fig. 10(a)** and **(b)**). The defect causes no significant contrast in the zero-order CBED spot meaning that the defect consists of only atomic position misalignments, which introduce a significant additional phase shift between the electron waves scattered from the two layers. Note that in diffraction imaging mode, where only diffraction peaks are detected, imaging of such a defect would not be possible. The area of overlap between CBED spots from the graphene and hBN layers is less in the higher order diffraction spots. The intensity contrast caused by the corrugation is more pronounced in the higher order CBED spots, in according with Eqs. (6) and (7). The CBED pattern was holographically reconstructed by the procedure described above. **Fig. 10(c)** and **(d)** show the recovered out-of-plane $\Delta z$ and in-plane $\Delta x$ atomic displacements. **Fig. 10(e)** compares the out-of-plane and in-plane atomic shifts along the ripple. The retrieved height of the out-of-plane ripple in hBN layer is about 2 nm, which agrees well with the observed out-of-plane ripples in graphene/hBN stacks due to self-cleansing effects [26].

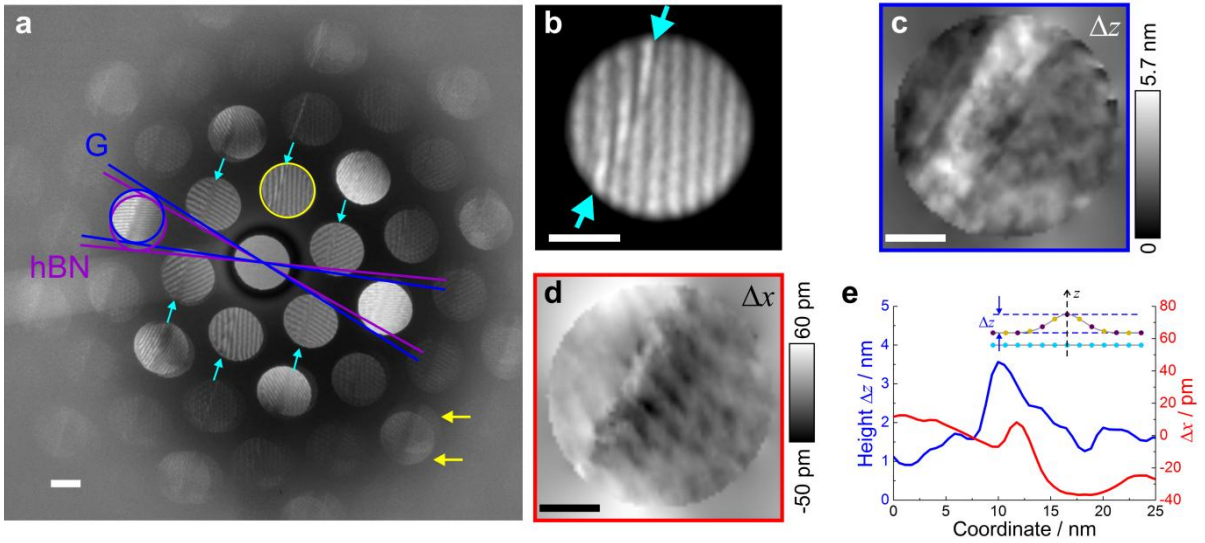

**Fig. 10.** Extracting the shape of an out-of-plane ripple from a CBED pattern. (a) Experimental CBED pattern acquired at defocus $\Delta f$ = -3 µm, with irregularity in the interference patterns marked by the arrows. The blue and purple lines indicate the relative rotation between graphene and hBN layers, which amounts to 3°. The cyan arrows indicate an out-of-plane ripple observed in the first-order CBED spots. The yellow arrows indicate the separation of CBED spots originating from graphene and hBN layers, where it becomes clear that the ripple is in the hBN layer. The intensity of the central spot is reduced by a factor of 0.1 for clarity. The scale bar corresponds to 2 nm$^{-1}$. (b) Magnified selected CBED spot (circled yellow in (a)) where irregularities of the fringe pattern can be seen. The scale bar corresponds to 1 nm$^{-1}$. (c) The reconstructed distribution of the ripple height $\Delta z$. (d) The reconstructed distribution of the lateral shift $\Delta x$. (e) Profiles for the magnitude of $\Delta z$ and $\Delta x$ profiles perpendicular to the ripple in (c) and (d). The scale bars in (c) and (d) correspond to 10 nm in real space. Adapted from [19].

### 4.3 Reconstruction of interlayer distance

HCBED can be applied for reconstruction of interlayer distances from a single CBED pattern as was demonstrated in reference [19] and explained in more details in reference [22]. The reconstruction procedure consists of the steps provided above, and the reconstructed phase distribution allows recovery of the interlayer distance according to the formula [22]

$$\Delta\varphi = \frac{\pi d}{\lambda} \tan\vartheta^{(2)} \tan\vartheta^{(1)} \cos\beta, \qquad (25)$$

where $\vartheta^{(1)}$ and $\vartheta^{(2)}$ are the diffraction angles of the two layers. For BLs where both layers are identical $\vartheta^{(1)} = \vartheta^{(2)} = \vartheta$ and for small twist angles, we obtain:

$$\Delta\varphi \approx \frac{\pi d \lambda}{a^2}. \tag{26}$$

**Fig. 11** shows the simulated 2D distributions and one-dimensional (1D) profiles of the reconstructed interlayer distances for BL graphene (BLG) with the interlayer distances of 0 and 10 Å at three different twist angles of 0.5°, 2° and 4°. From **Fig. 11**, we see that the reconstructed interlayer distances obtained from CBED patterns with smaller twist angles exhibit a smoother appearance, while at larger twist angles, artefact due to moiré structure become more pronounced in the reconstructions. The precision of the reconstructed interlayer distance is about ±0.5 Å [22].

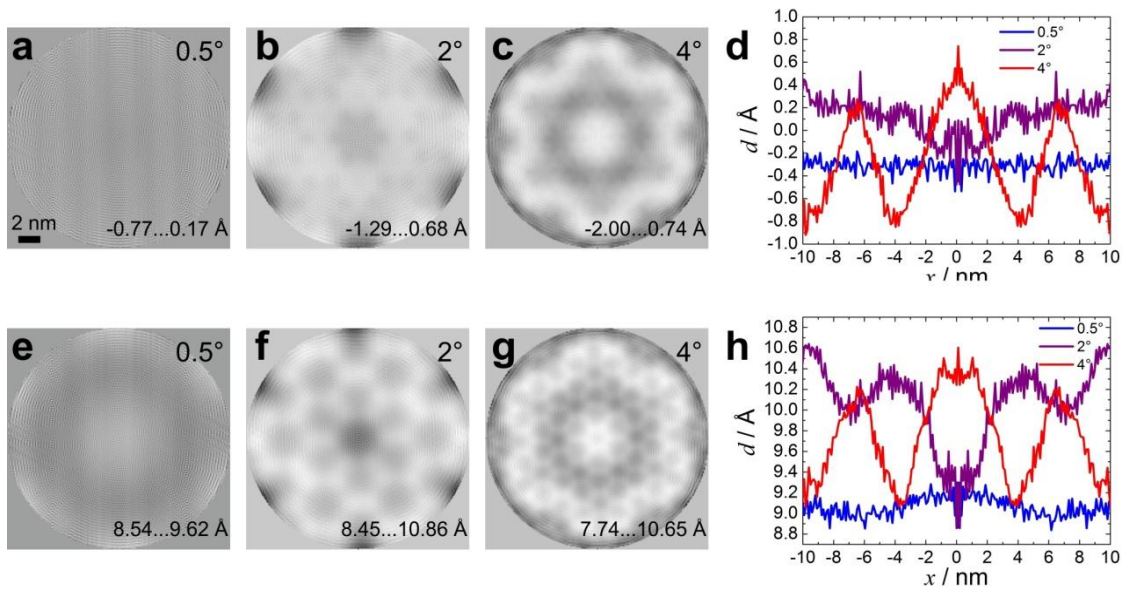

**Fig. 11**. 2D distributions ((a)-(c) and (e)-(g)) and 1D profiles through the middle of the 2D distributions ((d) and (h)) of the reconstructed interlayer distances from CBED patterns of BLG with the interlayer distances of 0 (a)-(d) and 10 Å (e)-(h), for three different twist angles of 0.5°, 2° and 4°. The vales given bottom right in (a)-(c) and (e)-(f) refer to the interlayer distance. Defocus distance is $\Delta f$ = -2 µm. Adapted from [22].

## 5. HCBED on multilayer systems, CBED moiré

Twisted multilayer systems exhibit extra modulations of the interference fringes in CBED patterns, i. e. a CBED moiré, as shown in **Fig. 12**. These extra modulations are coming from the interference between the moiré CBED spots. Such moiré CBED spots are the results of a second order process, which involves electrons being scattered by both layers sequentially [27]. Due to its second order nature such moiré peaks have much weaker intensity than the main diffraction peaks [27], and therefore the moiré CBED spots are also much weaker that the main CBED spots. This is the reason why they are practically invisible in twisted BL samples. Moreover, moiré CBED spots strongly overlap with the main CBED spots. For these two reasons, moiré CBED spots are not directly visible in the CBED patterns but they manifest themselves in the additional modulations of the CBED interference pattern - the CBED moiré, which is created by the interference between the moiré CBED spots and the major CBED spots.

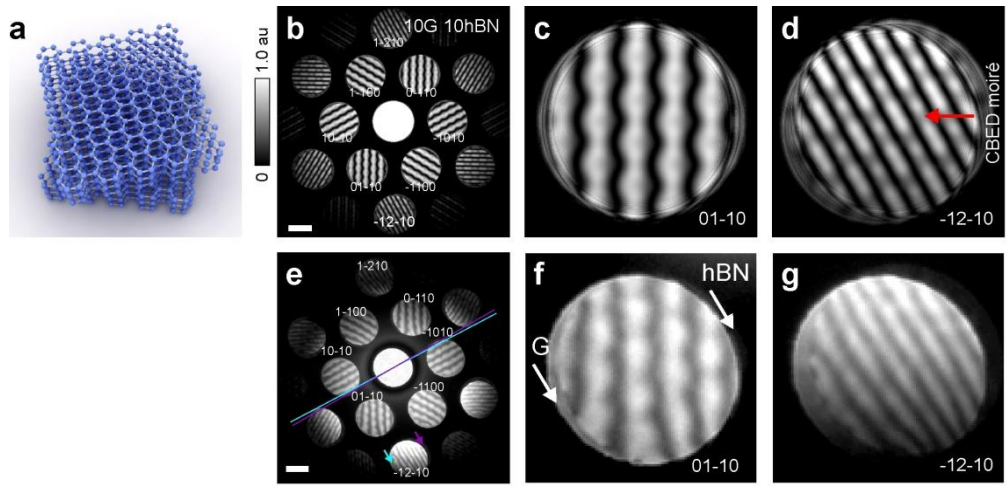

Fig. 12. CBED on twisted multilayer systems. (a) Illustration of twisted multilayer graphene. (b) Simulated CBED patterns for a multilayer van der Waals structure consisting of 10 graphene and 10 hBN layers, the twist angle between graphene and hBN layers is 2°, at a defocus, $|\Delta f|$ = 2.0 μm. (c) and (d) magnified CBED spots where a CBED moiré is observed in CBED interference fringes. (e) Experimental CBED patterns acquired at a defocus, $|\Delta f|$ = 2.0 μm, (f) and (g) magnified CBED spots. The scalebars in (b) and (e) are 2 nm$^{-1}$. Adapted from [21].

A simple and robust method for evaluation of the composition and the number of layers from a single-shot CBED pattern was demonstrated in reference [21]. The composition and the relative number of layers can be evaluated from the intensity distribution in the non-overlapping regions of a CBED spot. Number of layers can be evaluated by inspecting a 2D Fourier spectrum of a CBED spot, where the presence of peaks due to the CBED moiré indicate that there are five or more layers in the sample. Although the precision of such sample characterisation is very modest when compared to cross-sectional TEM imaging [28], the presented approach has the advantage that it is non destructive, requires only a single CBED pattern, and CBED is relatively easy to realise in a conventional TEM.

## 6. Discussion and outlook

HCBED offers a lot of information in one single CBED image such as atomic 3D displacements in the layers, the interlayer distances and imaging of adsorbates. Adsorbates can be clearly distinguished from 3D displacements of atoms: an adsorbate is identified in all orders of CBED spots, while atomic displacements produce intensity variations only in the first and higher order CBED spots. The contrast of the adsorbate image allows us to determine whether the adsorbate is an absorbing or phase object. Phase adsorbates display opposite intensity contrast when the probing wave is changed from convergent to a divergent wavefront (under focus to over focus). An absorbing adsorbate is displayed as a dark feature in both imaging regimes. The adsorbate distribution can be reconstructed from the zero-order CBED spot by applying an in-line hologram reconstruction routine. 3D displacements of atoms, on the other hand, display minimal intensity variations in the zero-order CBED spot, but they cause significant contrast in the higher-order CBED spots. Moreover, the type of displacement can be clearly identified just by comparing opposite CBED spots: an in-plane displacement leads to an opposite intensity contrast in opposite CBED spots (antisymmetric pattern), while an out-of-plane displacement leads to the same intensity contrast in the opposite CBED spots (pattern is symmetric).

For BL samples, the interference patterns formed in overlapping CBED spots can be treated as off-axis holograms and the phase of the interfering waves, and with this the 3D positions of the scattering atoms, can be retrieved. By using this approach, in-plane and out-of-plane ripples, and the interlayer distances can be quantitatively reconstructed. The resolution at which the atomic shifts are recovered exceeds the intrinsic resolution provided by the classical resolution criteria. The lateral and axial (along the $z$ axis) resolutions evaluated from a CBED pattern $k$-value range is given by $d_{x,y} = \frac{\lambda}{\sin \vartheta_{\max}}$ and $d_z = \frac{\lambda}{1 - \cos \vartheta_{\max}}$, respectively, where $\vartheta_{\max}$ is the maximal detected diffraction angle in the CBED pattern. According to these formulas, for a BL graphene CBED pattern acquired only up to the first-order CBED spots, the lateral resolution is $d_{x,y} = 2.13$ Å and the axial resolution is $d_z = 217.2$ Å. It is therefore a remarkable result that the holographic CBED approach allows reconstruction of interlayer distances of a few Angstroms at 0.5 Å accuracy, which is more than 400 times the diffraction defined $z$-resolution [19, 22].

From the experimental point of view, the best probing beam for acquiring CBED patterns would be a perfect convergent electron beam. However, holographic techniques in general, show very high tolerance to the probing beam imperfections, because the resulting interference pattern (hologram) is formed due the difference of the phases in the scattered and reference waves. In holographic CBED, the interference contrast is formed due to the relative phase shift between the waves scattered of atoms in different 2D layers. This relative phase shift is given by the local arrangement of the atoms in the structure, and the atoms are probed with the same local distribution of the probing beam. Thus, the probing beam imperfections, even when present, should have minimal effect on the CBED pattern formation, and with this, on the resulting reconstruction. However, a study can be performed to quantitatively evaluate the effect of beam aberrations, in particular, imperfections in the phase distribution, on the resulting reconstructions.

To conclude, HCBED has already demonstrated capability of providing high-resolution information about atomic arrangement from a single CBED pattern. We expect that HCBED can be further developed to become a practical tool for studying 2D materials at atomic resolution in 3D for complex heterostructures and arbitrary numbers of layers.

## Appendix: Simulation of CBED patterns

**Monolayer samples.** Simulation of CBED patterns was performed as follows. For calculation of the CBED pattern of a ML with displaced atoms, the input data is an array of coordinates of all atoms $(x_n, y_n, z_n)$. The far-field wavefront distribution of the scattered wave is simulated by summing up the waves scattered by each individual atom:

$$U(K_x, K_y) = \sum_n \psi_0(\vec{r}_n) \exp\left[-i(K_x x_n + K_y y_n)\right] \exp\left(-i z_n \sqrt{K^2 - K_x^2 - K_y^2}\right),$$

where $\psi_0(\vec{r}_n)$ is the complex-valued value of the probing wave at the *n*-th atom location, $\psi_0(\vec{r})$ is calculated by Eq. (5). No fast Fourier transforms are applied in the simulations to avoid sampling artifacts.

**Bilayer and multilayer samples.** For simulations of CBED patterns on multilayer samples it is assumed that all atoms in one layer have the same *z* position and the following simulation procedure is applied [22]. The transmission functions of MLs are calculated as:

$$t(x, y) = \exp\left[i\sigma v_z(x, y) \otimes l(x, y)\right], \tag{A1}$$

where $v_z(x, y)$ is the projected potential of an individual atom, $l(x, y)$ is the function describing positions of the atoms in the lattice, and $\otimes$ denotes convolution. The projected potential of a single carbon atom is simulated in the form:

$$v_z(r) = 4\pi^2 a_0 e \sum_{i=1}^{3} a_i K_0\left(2\pi r \sqrt{b_i}\right) + 2\pi a_0 e \sum_{n=1}^{3} \frac{c_i}{d_i} \exp\left(-\pi^2 r^2 / d_i\right),$$

where $r = \sqrt{x^2 + y^2}$, $a_0$ is the Bohr' radius, $e$ is the elementary charge, $K_0(...)$ is the modified Bessel function and $a_i, b_i, c_i, d_i$ are parameters that depend on the chemical origin of the atoms and are tabulated elsewhere [29]. In $v_z(r)$, the singularity at $r = 0$ is replaced by the value of $v_z(r)$ at $r$ = 0.1 Å. The convolution $v_z(x,y) \otimes l(x,y)$ in Eq. (A1) is calculated as $\text{FT}^{-1}\{\text{FT}[v_z(x,y)]\text{FT}[l(x,y)]\}$, where FT denotes Fourier transform. $\text{FT}[l(x,y)]$ is simulated without applying Fast Fourier transforms (FFT) to avoid artefacts associated with FFT, it is calculated as $\text{FT}[l(x,y)] = \sum_n \exp[-i(k_x x_n + k_y y_n)]$, where $(x_n, y_n)$ are the exact atomic coordinates, not pixels. The inverse FT is calculated by applying inverse FFT to the product of $\text{FT}[v_z(x,y)]$ and $\text{FT}[l(x,y)]$.

Each ML is assigned a transmission function calculated as described above. The incident convergent wave distribution $\psi_0(\vec{r})$ is calculated by simulating the diffraction of a spherical wavefront on an aperture (the second condenser aperture). No weak phase object approximation is applied in the simulations. The exit wave after passing through the first layer is given by the product of the incident wave and the transmission function of the first layer $u_1(x_1, y_1) = \psi_0(x_1, y_1) t_1(x_1, y_1)$. Next, this wave is propagated to the second layer. The propagation is calculated by the angular spectrum method [25]. The distribution of the propagated wave in the plane $(x_2, y_2)$ is $u_{2,0}(x_2, y_2)$. The exit wave after passing the second ML is calculated as $u_2(x_2, y_2) = u_{2,0}(x_2, y_2) t_2(x_2, y_2)$ and so forth. The CBED pattern is then simulated as the square of the amplitude of the FT of the exit wave after passing through the last layer.

## Acknowledgements

The authors acknowledge support from EU Graphene Flagship Program (contract CNECTICT-604391), European Research Council Synergy Grant Hetero2D (contract 319277), European Research Council Starting Grant EvoluTEM (contract 715502), the Royal Society, EPSRC grants EP/S019367/1, EP/P026850/1 and EP/N010345/1, EP/P009050/1, EP/S021531/1, FLAG-ERA project TRANS2DTMD.